\documentclass[journal]{IEEEtran}
\usepackage{cite}

\ifCLASSINFOpdf
\usepackage[pdftex]{graphicx}
\else
\fi
\usepackage{amsmath}
\ifCLASSOPTIONcompsoc
\usepackage[caption=false,font=normalsize,labelfont=sf,textfont=sf]{subfig}
\else
\usepackage[caption=false,font=footnotesize]{subfig}
\fi
\usepackage{url}

\newcommand{\ie}{i.e.~}
\newcommand{\eg}{e.g.~}

\usepackage{amssymb,amsthm,esint,wasysym}
\usepackage[colorlinks=true,linkcolor=black,anchorcolor=black,citecolor=black,filecolor=black,menucolor=black,runcolor=black,urlcolor=black]{hyperref}
\usepackage[all]{hypcap} 
\usepackage[capitalise]{cleveref} 
\usepackage{siunitx}
\DeclareSIUnit\sq{\ensuremath{\Box}}
\usepackage[version=4]{mhchem}
\usepackage{xcolor}

\crefname{equation}{}{}
\Crefname{equation}{Equation}{Equations}
\crefname{figure}{Fig.}{Fig.}
\Crefname{figure}{Fig.}{Fig.}

\usepackage{tikz}
\usepackage{textcomp}
\newcommand\copyrighttext{%
	\footnotesize \textcopyright 2021 IEEE. Personal use of this material is permitted.
	Permission from IEEE must be obtained for all other uses, in any current or future
	media, including reprinting/republishing this material for advertising or promotional
	purposes, creating new collective works, for resale or redistribution to servers or
	lists, or reuse of any copyrighted component of this work in other works.\\
	DOI: \href{https://doi.org/10.1109/TTHZ.2021.3095429}{10.1109/TTHZ.2021.3095429}}
\newcommand\copyrightnotice{%
	\begin{tikzpicture}[remember picture,overlay]
		\node[anchor=north,yshift=-5pt] at (current page.north)
		 {\fbox{\parbox{\dimexpr0.85\textwidth-\fboxsep-\fboxrule\relax}{\copyrighttext}}};
	\end{tikzpicture}%
}

\begin{document}
    \bstctlcite{IEEEexample:BSTcontrol}
	%
	\title{Terahertz Band-Pass Filters for Wideband Superconducting On-chip Filter-bank Spectrometers}
	%
	%
	\author{
		Alejandro~Pascual~Laguna,
		Kenichi~Karatsu,
		David~J.~Thoen,
		Vignesh~Murugesan,\\
		Bruno~T.~Buijtendorp,
		Akira~Endo,
		and~Jochem~J.~A.~Baselmans%
		\thanks{This work was supported by the European Research Council Consolidator Grant (ERC-2014-CoG MOSAIC), no. 648135, and in part by the Netherlands Organization for Scientific Research NWO under Vidi grant no. 639.042.423. (Corresponding author: Alejandro Pascual Laguna.)}%
		\thanks{A. Pascual Laguna, K. Karatsu, and J. J. A. Baselmans are with SRON Netherlands Institute for Space Research, 2333 CA Leiden, The Netherlands, and also with the THz Sensing group, Microelectronics Department, Delft University of Technology, 2628 CD Delft, The Netherlands (e-mail: A.Pascual.Laguna@sron.nl; K.Karatsu@sron.nl; J.Baselmans@sron.nl).}
		\thanks{V. Murugesan is with SRON Netherlands Institute for Space Research, 2333 CA Leiden, The Netherlands (e-mail: V.Murugesan@sron.nl).}
		\thanks{B. T. Buijtendorp is with the THz Sensing group, Faculty of Electrical Engineering, Mathematics and Computer Science, Delft University of Technology, 2628 CD Delft, The Netherlands (e-mail: B.T.Buijtendorp@tudelft.nl).}
		\thanks{D. J. Thoen and A. Endo are with the THz Sensing group, Faculty of Electrical Engineering, Mathematics and Computer Science, Delft University of Technology, 2628 CD Delft, The Netherlands, and also with the Kavli Institute of NanoScience, Delft University of Technology, 2628 CJ Delft, The Netherlands (e-mail: D.J.Thoen@tudelft.nl; A.Endo@tudelft.nl).}
	}

	\maketitle
	
	\begin{abstract}
		A superconducting microstrip half-wavelength resonator is proposed as a suitable band-pass filter for broadband moderate spectral resolution spectroscopy for terahertz (THz) astronomy. The proposed filter geometry has a free spectral range of an octave of bandwidth without introducing spurious resonances, reaches a high coupling efficiency in the pass-band and shows very high rejection in the stop-band to minimize reflections and cross-talk with other filters. A spectrally sparse prototype filter-bank in the band 300--400 GHz has been developed employing these filters as well as an equivalent circuit model to anticipate systematic errors. The fabricated chip has been characterized in terms of frequency response, reporting an average peak coupling efficiency of 27\% with an average spectral resolution of 940.
	\end{abstract}

	\copyrightnotice

	%
	\IEEEpeerreviewmaketitle
	
	\section{Introduction}
 	\label{sec:intro}
	
		\IEEEPARstart{N}{ext} generation imaging spectrometers for far-infrared astronomy \cite{highZgalaxies,submmGalaxies,THz_astronomy,CIB} will benefit from the development of compact on-chip terahertz (THz) filter-banks \cite{DESHIMA1,SuperSpec,CAMELS} with a spectral resolution matched to the typical emission linewidth of extra-galactic sources of $R\approx500$ \cite{highZgalaxies}. This advancement will allow the construction of true integral field units \cite{IFU}, which are crucial for the tomographic mapping of the early Universe \cite{3dsky_1,3dsky_2,SZ_astrophysics}, but are yet nonexistent in the far-infrared regime.
		
		The single-pixel on-chip THz filter-bank spectrometer DESHIMA is based upon NbTiN, a superconducting material with a critical temperature of $T_c\approx\SI{15}{\kelvin}$, which allows circuitry with negligible conductor loss up to the the gap frequency $f_\mathrm{gap}\approx\SI{1.1}{\tera\hertz}$. The former version of DESHIMA \cite{DESHIMA1}, despite showing technological readiness for astronomy at a ground-based telescope \cite{DESHIMA1_nature}, had several issues in its filters. The reported performance of the 49 channels sampling the \SIrange{332}{377}{\giga\hertz} spectrum averaged to a spectral resolution of $\langle{}R\rangle{}\approx300$ and a coupling efficiency of $8\%$ with respect to the signal entering the filter-bank. Both the low coupling efficiency and the low spectral resolution were caused by the radiative losses in its co-planar filters and the ohmic (conductive and dielectric) losses in the bridges balancing the potentials of the ground planes. Moreover, the filter design used was incompatible with broadband operation. In order to allow for truly wideband and highly-efficient filter-bank spectrometers a new band-pass filter design was required. As we shall see later in the manuscript, co-planar technology is not suitable for THz filter-banks because it incurs in radiation problems. For this reason we opted for microstrip technology to develop the new filters.
		
		In this paper we propose the design of a microstrip band-pass filter suitable for mid-resolution broadband THz on-chip spectrometers. We target filters with a coupling efficiency to the detectors as high as possible, an octave of instantaneous bandwidth free from spurious resonances and a moderate spectral resolution of $R=500$. The rest of the article is organized as follows. In \cref{section:FB}, we revisit the concept of the superconducting on-chip filter-bank spectrometer. In \cref{sec:resAsBandPassFilter} we investigate the foundations of resonators as band-pass filters and construct a circuit model, largely inspired by \cite{SuperSpecCircuitModel,GeorgeChePhD}, to study large filter-banks. In \cref{sec:chipTech}, we explain why co-planar waveguide (CPW) technology should not be used for THz filter-banks and motivate the use of microstrips instead. In \cref{sec:filterDesign}, a microstrip band-pass filter geometry is proposed and used to design a filter-bank sparsely sampling the spectrum \SIrange{300}{400}{\giga\hertz}, whose fabrication details and measurements are discussed in \cref{sec:sparseFB}. The concluding remarks are outlined in \cref{sec:conclusions}.
		
		\IEEEpubidadjcol

	\section{On-Chip Filter-Bank Spectrometer}
	\label{section:FB}
\begin{figure}[t!]
			\centering
			\includegraphics{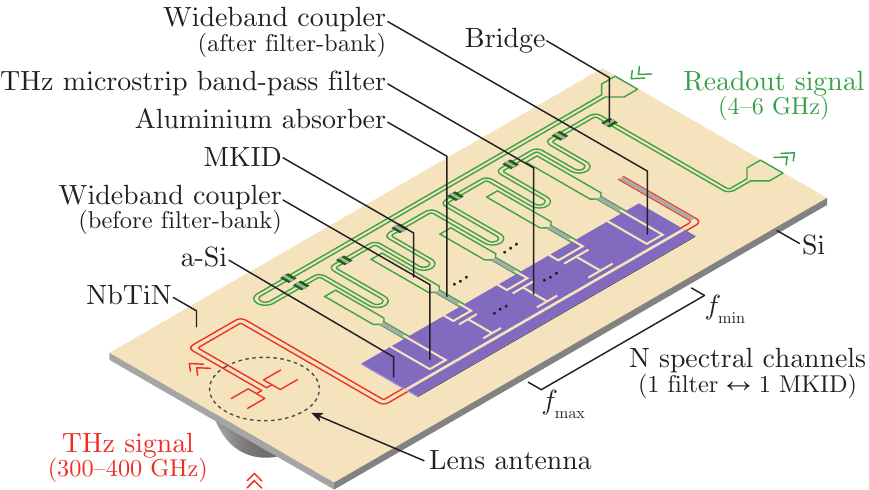}
			\caption{Conceptual drawing of a superconducting on-chip filter-bank spectrometer based on MKIDs. Note the different filter technology and shape with respect to \cite{DESHIMA1}.}
			\label{fig:chip_cartoon}
		\end{figure}

		In this paper, we specifically target a fully sampled superconducting on-chip THz filter-bank spectrometer with a constant spectral resolution $R=500$. Such a spectrometer couples THz radiation to the chip with a broadband antenna, which transforms the incoming radiation onto a guided mode. From the antenna, a superconducting transmission line, hereinafter called through-line, guides the wideband signal with virtually no loss of power \cite{radiationLossCPW} to the filter-bank, where it is sorted into sub-bands and subsequently detected. Any remaining power not detected by the filter-bank is eventually absorbed in a matched lossy transmission line terminating the through-line to avoid reflections. Before and after the filter-bank there are a few weakly coupled detectors which, thanks to the their wideband response, serve for calibration and diagnostic purposes. A schematic drawing of such a device is depicted in \cref{fig:chip_cartoon} and described in great detail in \cite{DESHIMA1}.
		
		Each of the $N$ band-pass filters in the filter-bank is tuned to channelize a sub-band of the broadband THz signal from the through-line to the detector, where the filtered signal is sensed and read out. Since the aim is to fully sample the whole operational bandwidth ${f_\mathrm{min}:f_\mathrm{max}}$ with a constant spectral resolution ${R=f_i/\delta{}f_i}$ and response cross-overs at \SI{-3}{\decibel}, each filter has an exponentially lower central frequency ($f_i$) and a narrower \SI{-3}{\decibel} pass-band ($\delta{}f_i$) following the relation ${f_i=f_\mathrm{max}(1+R^{-1})^{-i+1}}$ for the $i^\mathrm{th}$ channel from the highest frequency, which is defined as ${f_\mathrm{max}=f_\mathrm{min}(1+R^{-1})^{N-1}}$.
		
		Every sub-band is sensed with a hybrid Microwave Kinetic Inductance Detector (MKID) \cite{MKID,hybridMKID}, which uses a small aluminium strip to absorb the radiation coupled through each filter at frequencies beyond its gap frequency $f_\mathrm{gap}\approx\SI{90}{\giga\hertz}$. By capacitively coupling an array of MKIDs with slightly different lengths to a single readout line, each detector can be simultaneously sensed at a different frequency with a comb of microwave probing tones, allowing a highly frequency-multiplexed readout scheme \cite{KIDreadout_Baselmans,KIDreadout_Mazin}.

	\section{Resonator as a THz Band-pass Filter}
	\label{sec:resAsBandPassFilter}
	
		Each shunted band-pass filter should maximize the coupling efficiency in the pass-band, \ie the power transferred from the through-line into the MKID; and in the stop-band it should not affect the THz signal on the through-line. The ideal top-hat band-pass response could be approximated with multiple filtering elements \cite{coupledFiltersTopHat,oversamplingFB}, however to limit the chip complexity, this is not done. Moreover broadband lumped-element solutions like \cite{cochlea} become very complex at THz frequencies due to their electrical length. Thereby, we have investigated resonators as a simple broadband distributed band-pass filter solution, as these naturally provide an octave of spurious-free bandwidth for a half-wavelength resonator. The bandwidth of the Lorentzian pass-band of these resonators is controlled by means of the coupling strength of the surrounding couplers.
		
\begin{figure}[t!]
			\centering
			\includegraphics{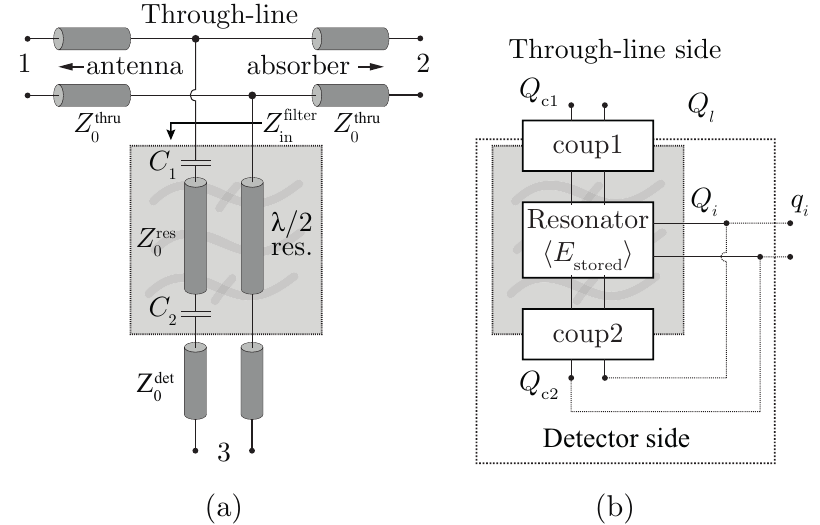}
			\caption{A resonator as a band-pass filter from a circuit point of view, and (b) an energetic point of view.}
			\label{fig:circuit_and_resonator_pictures}
		\end{figure}

\begin{figure}[t!]
			\centering
			\includegraphics{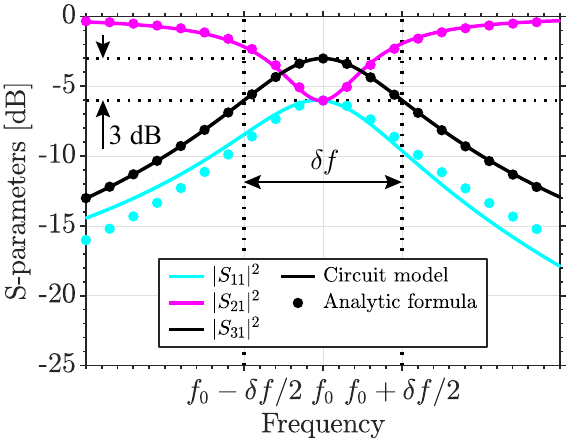}
			\caption{Frequency response of an isolated lossless band-pass filter with spectral resolution $R$ as in \cref{fig:circuit_and_resonator_pictures}. The analytical expressions \cref{eq:S11_f0,eq:S21_f0,eq:S31_f0} are shown with dots and the results of the numerical circuit model are represented with solid lines.}
			\label{fig:isolatedFilterResponse}
		\end{figure}

		By defining the ports of a single band-pass filter coupled to a through-line as in \cref{fig:circuit_and_resonator_pictures}(a), the figure of merit to maximize on-resonance is the coupling efficiency or $|S_{31}|^2$; and off-resonance the transmission coefficient $|S_{21}|^2$, so that most of the power is extracted on-resonance and off-resonance the filter does not load the through-line. It turns out that for a lossless filter, if no extra filtering structures like band-stop filters in the through-line are added \cite{WSPEC,Matthei_decouplingResonators}, at most 50\% of the input power can be extracted on-resonance (${|S_{31}|^2=50\%}$). The remaining 50\% of the input power is evenly split between reflections (${|S_{11}|^2=25\%}$) and power let to continue in the through-line (${|S_{21}|^2=25\%}$). This frequency response is illustrated in \cref{fig:isolatedFilterResponse}. In the following we shall see this band-pass filter in two different ways: as a distributed circuit model (\cref{fig:circuit_and_resonator_pictures}(a)), and as a resonator with several energy-leaking mechanisms (\cref{fig:circuit_and_resonator_pictures}(b)).
		
		\subsection{Resonator Perspective}
		\label{sec:resonatorModel}
		
		The quality factor $Q$ of a resonator relates the average energy stored in it $\langle{}E_\mathrm{stored}\rangle{}$ to the power lost $P_\mathrm{lost}$ by the energy-leaking mechanisms at its resonant frequency $f_0$ by the expression
		\begin{equation}
			Q=2\pi{}f_0\frac{\langle{}E_\mathrm{stored}\rangle{}}{P_\mathrm{lost}}=\frac{f_0}{\delta{}f}.
			\label{eq:Qfactor}
		\end{equation}
		For the resonator of \cref{fig:circuit_and_resonator_pictures}(b), the energy-leaking mechanisms are two couplers and any internal loss mechanism in the resonator. One coupler leaks energy from the resonator to the through-line with an associated quality factor $Q_{c1}$, and the other coupler to the detector with $Q_{c2}$. The loss mechanisms, represented by $Q_i$, will be radiative or dielectric losses for a superconducting device. As a result, the loaded quality factor $Q_l$ of the resonator can be written as
		\begin{equation}
			Q_l^{-1}=Q_{c1}^{-1}+Q_{c2}^{-1}+Q_{i}^{-1}.
			\label{eq:Ql}
		\end{equation}
		
		Following a similar approach as in \cite{BenMazinPhD,Akira_APL}, the 3-port network S-parameters of the shunted half-wavelength resonator of \cref{fig:circuit_and_resonator_pictures} may be approximated for frequencies $f$ around the resonance $f_0$ by means of the quality factors of its components. The dots in \cref{fig:isolatedFilterResponse} illustrate the equations of
		\begin{align}
			S_{11}(f)&\approx{}\frac{S_{11}(f_0)}{1+j2Q_l\frac{f-f_0}{f_0}}, \label{eq:S11_around_f0}\\
			S_{21}(f)&\approx{}\frac{S_{21}(f_0) + j2Q_l\frac{f-f_0}{f_0}}{1+j2Q_l\frac{f-f_0}{f_0}}, \label{eq:S21_around_f0}\\
			S_{31}(f)&\approx{}\frac{S_{31}(f_0)}{1+j2Q_l\frac{f-f_0}{f_0}}, \label{eq:S31_around_f0}
		\end{align}
		being $S_{11}(f_0)$, $S_{21}(f_0)$ and $S_{31}(f_0)$ the values of the S\nobreakdash-parameters on-resonance; which in turn are given by
		\begin{align}
			S_{11}(f_0)&=\frac{-q_i}{Q_{c1}+q_i}=\frac{-Q_{l}}{Q_{c1}}, \label{eq:S11_f0}\\
			S_{21}(f_0)&=\frac{Q_{c1}}{Q_{c1}+q_i}=\frac{Q_{l}}{q_i}, \label{eq:S21_f0}\\
			S_{31}(f_0)&=\frac{\sqrt{2Q_{c1}Q_{c2}}}{Q_{c2}+Q_{c1}\left(1+\frac{Q_{c2}}{Q_i}\right)},\label{eq:S31_f0}
		\end{align}
		where $q_i$ is the internal quality factor of the equivalent two-port network depicted in \cref{fig:circuit_and_resonator_pictures} and it is defined as
		\begin{equation}
			q_i=\left(Q_i^{-1}+Q_{c2}^{-1}\right)^{-1}.
			\label{eq:qi}
		\end{equation}
		This allows to re-write \cref{eq:Ql} as
		\begin{equation}
			Q_l^{-1}=Q_{c1}^{-1}+q_i^{-1}.
			\label{eq:Ql_2port}
		\end{equation}
		
\begin{figure}[t!]
			\centering
			\includegraphics{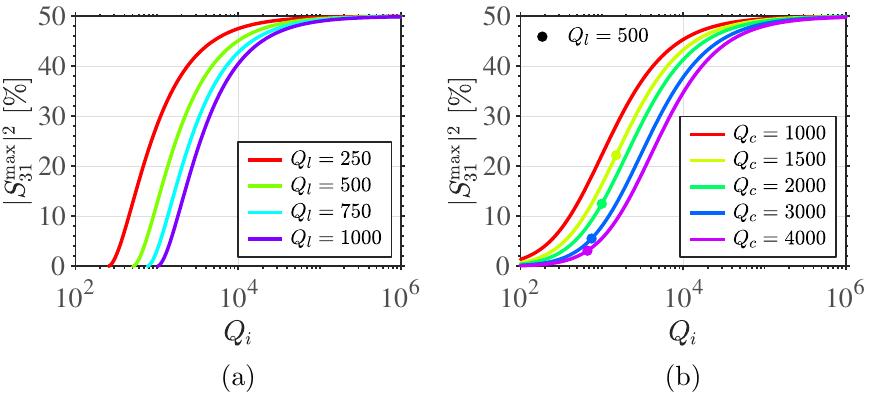}
			\caption{Maximum coupling efficiency as a function of the internal quality factor $Q_i$ for (a) several cases of loaded quality factor $Q_l$ and (b) several cases of coupling quality factor $Q_c=Q_{c1}=Q_{c2}$, with the location of $Q_l=500$ emphasized with dots.}
			\label{fig:S31max}
		\end{figure}

		By equating to zero the derivatives of \cref{eq:S31_f0} with respect to $Q_{c1}$ and $Q_{c2}$, one can find that the configuration that maximizes the coupling efficiency on-resonance, for a desired loaded quality factor $Q_l$ and a given internal quality factor $Q_i$, is when the two couplers have the same $Q_c$ with a value of
		\begin{equation}
			Q_c=Q_{c1}=Q_{c2}=\frac{2Q_iQ_l}{Q_i-Q_l}.
			\label{eq:optimumQc}
		\end{equation}
		Under this condition the coupling efficiency in \cref{eq:S31_f0} is maximized and peaks at
		\begin{equation}
			|S_{31}^\mathrm{max}(f_0)|^2=\frac{\left(Q_i-Q_l\right)^2}{2Q_i^2}=\frac{2Q_l^2}{Q_c^2}=\frac{2Q_i^2}{(2Q_i+Q_c)^2},
			\label{eq:optimumS31}
		\end{equation}
		which is represented in \cref{fig:S31max} as a function of the internal quality factor $Q_i$ for different values of $Q_c=Q_{c1}=Q_{c2}$. It becomes apparent that to obtain a high coupling efficiency ($|S_{31}(f_0)|^2$), the internal quality factor must be much larger than the targeted loaded quality factor ($Q_i\gg{}Q_l$).

		\subsection{Circuit Perspective}
		\label{sec:circuitModel}
		On the other hand, an equivalent circuit representation as in \cref{fig:circuit_and_resonator_pictures}(a) also allows us to investigate these resonators with simple microwave analysis techniques. The resonator is represented by a transmission line whose losses are characterized by $Q_i=\beta/(2\alpha)$, relating the real and imaginary parts of the propagation constant $\gamma=\alpha+j\beta$. The length of the resonator is determined by the resonance condition $\Im\{Z_\mathrm{in}^\mathrm{filter}\}=0$, which gives a slightly shorter length than $\lambda/2$ due to the detuning introduced by the couplers. The couplers surrounding the half-wavelength resonator are modeled by capacitors, which are designed to provide the coupling quality factor of \cref{eq:optimumQc} following an energetic approach as in \cite{RamiBarendsPhD} with $Q_{c}={2\pi}/{|S_{ab}|^2}$, where $|S_{ab}|^2$ is the transmission through a series capacitor between ports $a$ and $b$. This transmission can be calculated for a series capacitance $C$ using the following expression relating the S and ABCD parameters \cite{Frickey}:
		\begin{equation}
			S_{ij}=\frac{2\sqrt{\Re\{Z_0^a\}\Re\{Z_0^b\}}}{Z_0^a+(j\omega{}C)^{-1}+Z_0^b},
			\label{eq:S21_seriesCap}
		\end{equation}
		where $Z_0^a$ and $Z_0^b$ are the normalizing impedances at ports $a$ and $b$, respectively. The capacitor at the through-line side has $Z_0^a=Z_0^\mathrm{thru}/2$ and $Z_0^b=Z_0^\mathrm{res}$, whereas the capacitor at the detector side has $Z_0^a=Z_0^\mathrm{det}$ and $Z_0^b=Z_0^\mathrm{res}$.
		
\begin{figure}[t]
			\centering
			\includegraphics{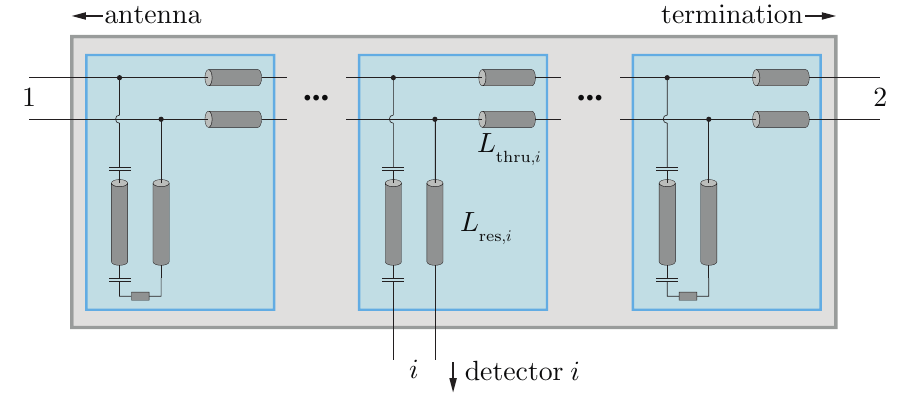}
			\caption{Filter-bank circuit model using the unit cell of \cref{fig:circuit_and_resonator_pictures} for each channel.}
			\label{fig:FBmodel}
		\end{figure}

		From these considerations the unit cell of the filter-bank, the band-pass filter of \cref{fig:circuit_and_resonator_pictures}(a), can be fully described in transmission line terminology for a given resonance $f_0$ and a given resolution $R=Q_l=f_0/\delta{}f$. By cascading a network comprised of these filters branching off a transmission line, a full filter-bank can be analyzed with a circuit model as depicted in \cref{fig:FBmodel} to obtain the S-parameters relating every port: input (1), detectors ($i$) and termination (2) of the filter-bank. To ease the calculations, an ABCD matrix approach was employed to link each pair of ports while the other were kept loaded. The S-parameters of the whole filter-bank can be easily obtained by means of the transformations in \cite{Frickey}.
		
		We analyze the response of a filter-bank fully sampling the octave band $220$--$440$ GHz with $N=347$ band-pass filters designed to have a spectral resolution in isolation of $R=Q_l=500$. Losses, characterized by $Q_i=3300$, are also accounted for in every transmission line. To minimize them, the filters are monotonically ordered along the through-line from the highest frequency ($f_\mathrm{max}$) to the lowest ($f_\mathrm{min}$), starting from the antenna side. The separation between contiguous filters along the through-line $L_{\mathrm{thru},i}$ is chosen to be $\lambda_{\mathrm{thru},i}/4$, being $\lambda_{\mathrm{thru},i}$ the effective wavelength of the microstrip mode propagating in the through-line at the resonance of the $i^\mathrm{th}$ filter. This distance must be kept in order to prevent additional reflections from appearing within an octave band, which would degrade the performance of a broadband filter-bank.
		
		\begin{figure*}[t]
			\centering
			\includegraphics{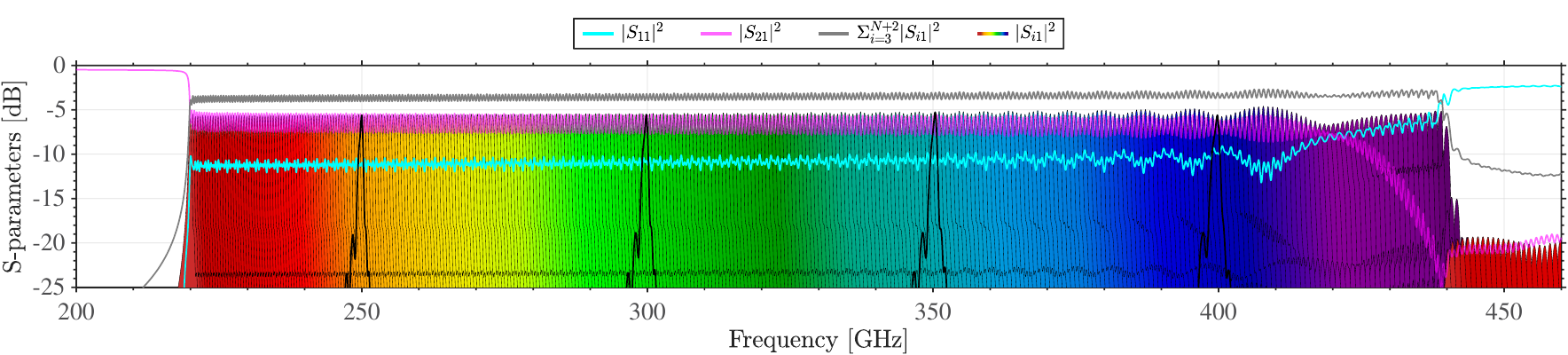}
			\caption{Simulated response for a filter-bank circuit model (\cref{fig:FBmodel}) fully sampling the octave band \SIrange{220}{440}{\giga\hertz} with $N=347$ filters ordered $f_\mathrm{max}\rightarrow{}f_\mathrm{min}$, with a targeted spectral resolution of $R=Q_l=500$, and an inter-filter separation of $L_{\mathrm{thru},i}=\lambda_{\mathrm{thru},i}/4$. The simulated loss corresponds to $Q_i=3300$, resulting in a maximum coupling efficiency of $-4.4$ dB for an isolated filter following \cref{eq:optimumS31}. Neighboring filters cross-talk and as result the performance of the filter-bank is different with respect to the isolated filters: the average peak coupling strength is approximately $-5.6$ dB and the average loaded quality factor around $422$. The pass-bands of four filters at around 250, 300, 350 and 400 GHz have been emphasized for clarity.}
			\label{fig:fullFilterBankResponse}
		\end{figure*}
		
		The S-parameters for the filter-bank just described are depicted in \cref{fig:fullFilterBankResponse}. As can be seen, the performance of the different channels of the filter-bank is similar to that of an isolated filter with a few differences: firstly, the average peak transmission in \cref{fig:fullFilterBankResponse} is $\langle|S_{i1}(f_i)|^2\rangle\approx{}-5.6$ dB (instead of $-4.4$ dB as expected from \cref{eq:optimumS31} for an isolated	filter with $Q_i=3300$) due to the overlapping pass-bands of the filters; and secondly, the leading filters of the filter-bank (higher frequencies) get more power due to the coherent addition of the reflections from the lower frequency filters in the proximity to their second harmonic. The reflection level at the input of the filter-bank ($|S_{11}|^2$) remains below $-6$ dB until the second harmonics of the low frequency resonators ring at frequencies higher than \SI{440}{\giga\hertz}. The fraction of the input power transferred to the absorber terminating the through-line ($|S_{21}|^2$) is high for frequencies below the operational band of the filter-bank and reduces as the different filters extract power from the through-line. An indication of how much of the input power is extracted by all the filters combined can be calculated using $\sum_{i=3}^{N+2}|S_{i1}|^2$, which averages to $-3.5$ dB in the band of operation in \cref{fig:fullFilterBankResponse}. The rest of the power is reflected at the entrance of the filter-bank, absorbed in its termination or associated to losses. The loaded quality factor averages $422$, which is lower than the intended value of 500 due to the interaction of the filters.

	\section{On-Chip Technology: CPW vs.~Microstrip}
	\label{sec:chipTech}
	
		In this section we delve into the choice of on-chip technology, as this decision carries important consequences for the overall performance of the filter-bank. First of all, because the metals used are superconductors, conductor losses are negligible \cite{MBtheory}. Dielectric losses, on the other hand, depend on whether the dielectrics in the proximity of the transmission lines are crystalline, \eg crystalline Si, or amorphous, \eg deposited dielectrics like a-Si \cite{Bruno}. Lastly, radiation losses largely depend on the technology choice and the overall size of the transmission line.
		
\begin{figure*}[t]
			\centering
			\includegraphics{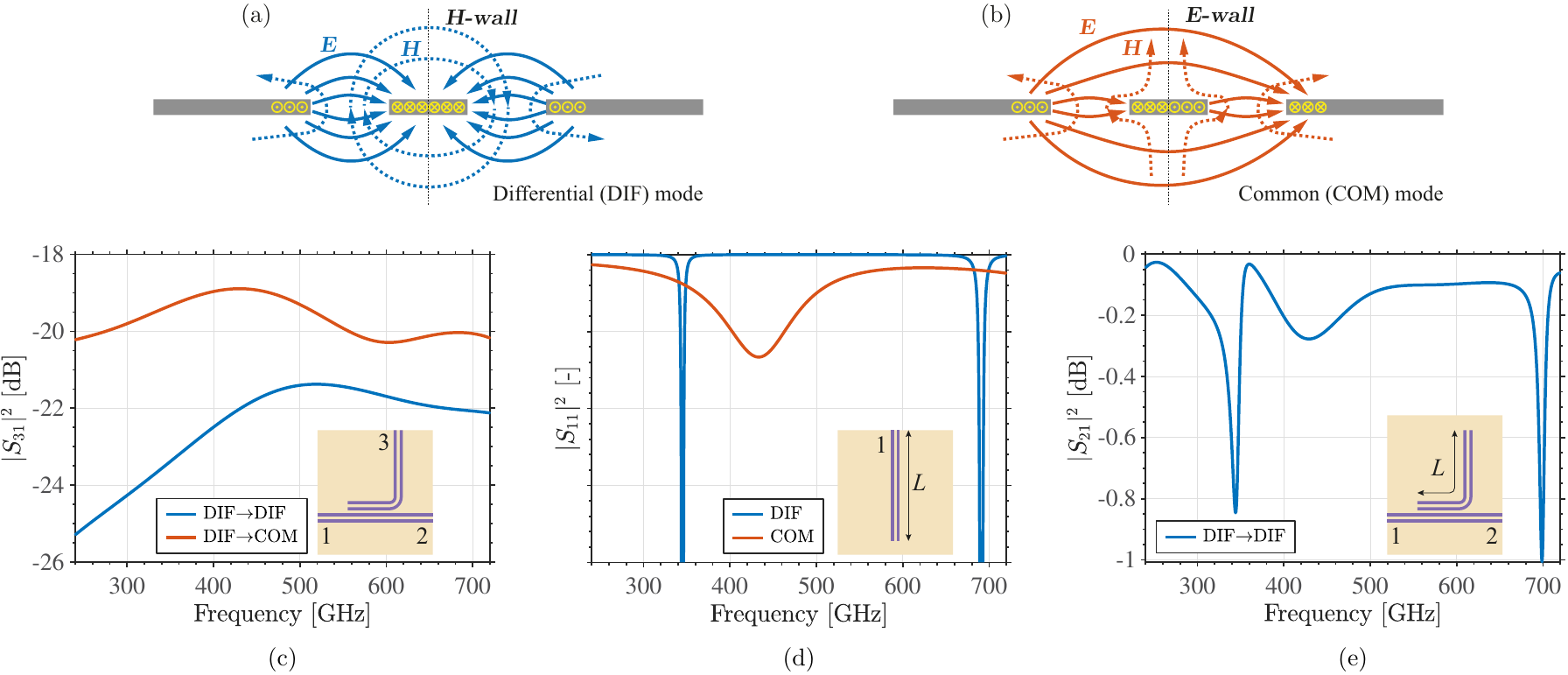}
			\caption{Sub-figures (a) and (b) show the two fundamental modes of a CPW: respectively differential (DIF) mode, and common (COM) mode. Sub-figure (c) shows the mode conversion from a differentially-excited CPW to an ``elbow''-coupled CPW. Sub-figure (d) shows the large difference in the dips associated to a straight CPW half-wavelength resonator excited with DIF mode or COM mode at port 1 with an arbitrarily small port impedance to emulate a short-circuit. The COM mode shows a much broader and shallower resonance than the DIF mode, which is associated to higher losses (lower $Q_i$). Lastly, sub-figure (e) shows the response of a differentially-fed CPW structure consisting of a through-line and a resonator. We observe sharp and broad dips similar to the ones associated to the two different modes in sub-figure (d), which indicates that the CPW resonator allows both modes to be excited. The DIF resonance is largely deteriorated by the COM mode dip. Moreover, the COM mode spoils the transmission for subsequent filters in the through-line.}
			\label{fig:CPW}
		\end{figure*}

		Despite the relatively easy manufacturing of CPW technology, this type of transmission line is problematic to work with at THz frequencies due to its support for two fundamental modes: the differential mode, for which the electric fields on the two slots have opposite directions (\cref{fig:CPW}(a)), and the common mode, with electric fields in the two slots oriented in the same direction (\cref{fig:CPW}(b)). The differential CPW mode is the weakly-radiative fundamental transmission line mode of this structure; \eg a differentially-excited superconducting CPW with a total width of \SI{6}{\micro\metre} has shown a very low radiation loss ($Q_i\approx15000$) \cite{radiationLossCPW}. On the contrary, the common mode is mostly a broadband radiative mode associated to a leaky wave when the slots are placed between two media with different permittivity \cite{slotTwoMedia}.
		
		A combination of these two fundamental modes will exist with any asymmetry around a CPW, and will thereby incur excess radiation loss at high frequencies (resulting in a low $Q_i$) \cite{TLTool,CPWmodes,DESHIMA1}. Balancing the potential of the grounds with bridges will only reflect the common mode, but it will not convert it to a differential mode. As a result, to minimize radiation loss in CPW structures at high frequencies, symmetry must be preserved. However, because the filter-bank concept requires to channelize the energy into parallel detectors, the symmetry will necessarily be broken in the through-line and the filters.
		
		To explicitly show the mode-conversion of a CPW-based tank circuit at high frequencies we simulate in CST Microwave Studio \cite{CST} the typical ``elbow'' coupler, extensively used for MKID designs \cite{elbowCoupler}, as a THz half-wavelength resonator. The simulations consist of CPW structures on an infinitely thin superconductor with a sheet inductance of \SI{1}{\pico\henry\per\sq} on top of a Si substrate. In \cref{fig:CPW}(c) we observe that the differential mode in the through-line actually couples more strongly to the common mode than to the differential mode in the coupled line. \Cref{fig:CPW}(d) shows that a resonator excited with a common mode will have in comparison with a differentially-excited resonator: a much broader resonance, due to the stronger coupling and the larger energy leakage (as seen from \cref{eq:Ql_2port}); and a shallower dip, due to the excess radiation losses (appreciated in \cref{eq:S21_f0}). Lastly, in \cref{fig:CPW}(e) we observe the same two modes of \cref{fig:CPW}(d) get intertwined in the response of a differentially-fed through-line with an ``elbow''-coupled resonator. This is detrimental for both the differential mode resonance and for prospective neighboring filters using the degraded transmission around the common mode resonance. Furthermore, a poor off-resonance transmission will also occur due to the use of a CPW through-line as in \cite{DESHIMA1}, whose asymmetric loading with filters will induce a common mode on it and will thereby largely deteriorate the overall filter-bank performance with excess radiation losses. As stated earlier, CPW bridges will not transform the common mode to a differential mode but rather reflect it, being eventually radiated.

	On the other hand, microstrip lines only support a single well-confined fundamental mode whose radiation losses are negligible \cite{TLTool}. However, practical microstrip devices require deposited dielectrics, which incur more losses than crystalline substrates as reported in \cite{microwaveLosses} for microwave frequencies. At THz frequencies data on dielectric loss is very sparse: a CPW on crystalline Sapphire shows a $Q_i$ in excess of $15000$ \cite{radiationLossCPW}, whereas microstrips fabricated from NbTiN and Plasma-Enhanced Chemical Vapor Deposition (PECVD) SiN report $Q_i\approx1400$ at \SI{200}{\giga\hertz} \cite{Superspec_Qi}, and NbTiN microstrips on PECVD a-Si \cite{Bruno} show $Q_i\approx4750$ at \SI{350}{\giga\hertz} \cite{microstripFP}. Despite the dielectric losses in deposited a-Si still being high for an ideal THz filter-bank, these are sufficient to design and build microstrip-based filters, thereby avoiding the mode-related issues of co-planar technology.

	\section{Band-Bass Filter Design}
	\label{sec:filterDesign}
		
		\begin{figure*}[t!]
			\centering
			\includegraphics{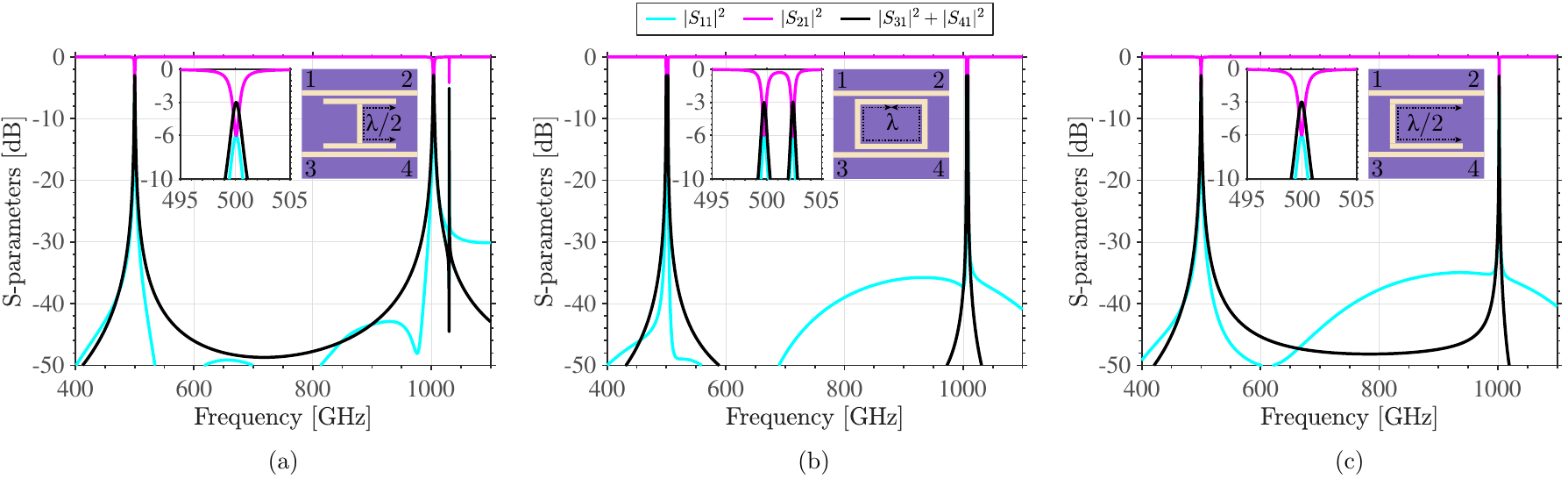}
			\caption{Comparison of the S-parameters simulated in Sonnet \cite{Sonnet} of three microstrip resonator geometries: (a) an I-shaped resonator, (b) a O-shaped resonator and (c) an C-shaped resonator. The three geometries show optimal performance on-resonance with $Q_l=500$, however, the O-shaped resonator shows a spurious resonance close to the intended one. The C-shaped resonator works well overall but the I-shaped resonator has a much better off-resonance performance. The I-shaped resonator has a high-Q spurious resonance associated to the coupling bar, but this falls beyond the octave band of interest.}
			\label{fig:comparisonResonators}
		\end{figure*}
		
		In this section we design a superconducting I-shaped microstrip resonator to work as a THz band-pass filter. In \cref{fig:comparisonResonators}, this geometry and its superior performance are compared against other structures such as the O-shaped and the C-shaped resonators. For all the three geometries, the microstrip layer is made of a thin NbTiN superconductor film with a sheet kinetic inductance of $L_k=\SI{1}{\pico\henry\per\sq}$. The microstrip sits on \SI{300}{\nano\meter} of a-Si ($\varepsilon_r=10$) and it is backed by a NbTiN ground plane with $L_k=\SI{0.448}{\pico\henry\per\sq}$. The resonators are parallel-coupled to the surrounding microstrip lines through \SI{300}{\nano\meter} gaps. The resonator microstrip lines are \SI{450}{\nano\meter} wide, with the exception of the vertical bar of the ``I'' resonator, which is \SI{1.1}{\micro\meter}.
		
		These three geometries have been simulated in Sonnet \cite{Sonnet} and their S-parameters in \cref{fig:comparisonResonators} show optimal on-resonance performance at \SI{500}{\giga\hertz} as described in \cref{sec:resAsBandPassFilter} and a loaded quality factor $Q_l\approx500$. The second harmonic of the resonators rings at twice the frequency, at \SI{1}{\tera\hertz}. The O-shaped resonator resonates when its perimeter becomes one wavelength, but it also shows a spurious resonance at a slightly lower frequency than the intended one associated to a mode with current nodes in the middle of the coupling bars \cite{spuriousResRing}. The spurious mode of the O-shaped resonator is in fact the fundamental resonant mode of the C-shaped half-wavelength resonator. This one, on the other hand, shows no spurious resonance and it has been successfully used for the filter-banks in \cite{SuperSpec,Akira_APL}. Lastly, the I-shaped resonator shows the best off-resonance performance with the lowest reflections ($|S_{11}|^2$) of the three resonators. This filter rings when the overall length measured between one coupler extreme to another extreme of the other coupler becomes half-wavelength as shown in the inset of \cref{fig:comparisonResonators}(a). Despite having a spurious resonance ($\sim\SI{1030}{\giga\hertz}$) associated to the resonance of the coupling bars, this could be tuned to fall beyond the intended octave free spectral range. Thereby we choose the I-shaped resonator option as it shows the best overall performance.
		
\begin{figure}[t!]
			\centering
			\includegraphics{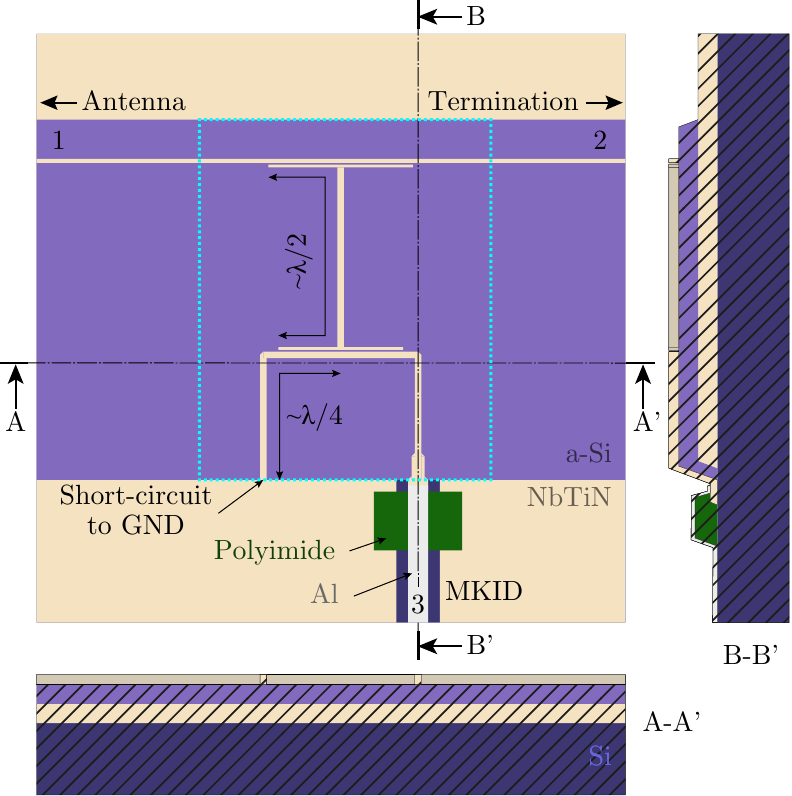}
			\caption{Views of the I-shaped microstrip band-pass filter. The thicknesses in the cross-sectional views A-A' and B-B' have been scaled by 150\% to emphasize the layering (except for the Si wafer, which is not to scale). A slanted hatch is displayed over the parts cut.}
			\label{fig:IshapedFilter}
		\end{figure}

		The actual implementation of the I-shaped resonator is depicted in \cref{fig:IshapedFilter}, which deviates slightly from the inset of \cref{fig:comparisonResonators}(c) with port 3 being terminated in a short-circuit to ground at $\lambda/4$ from the coupler and port 4 becoming port 3 in \cref{fig:IshapedFilter} as the only output to the detector. This filter comprises two couplers (horizontal bars of the ``I''), which are narrow microstrip lines (\SI{450}{\nano\meter} wide) coupling the resonator through a \SI{300}{\nano\meter} gap to either the through-line or the detector line, controlling respectively $Q_{c1}$ and $Q_{c2}$ with their length. The \SI{1.1}{\micro\meter} wide microstrip connecting the two couplers (vertical bar of the ``I'') was designed to have half the characteristic impedance of that of the coupler microstrip lines, resulting in an overall electrical length of the resonator of $\sim\lambda/2$, measured from one coupler extreme, through the vertical bar of the ``I'', to another extreme of the other coupler. The microstrip line on the detector side is short-circuited to ground at roughly $\lambda/4$ from the coupler in order to coherently reflect the filtered THz signal to the MKID while providing with a low electric field at the microwave frequencies of the readout to minimize TLS noise contributions \cite{TLS}.
		
\begin{figure}[t!]
			\centering
			\includegraphics{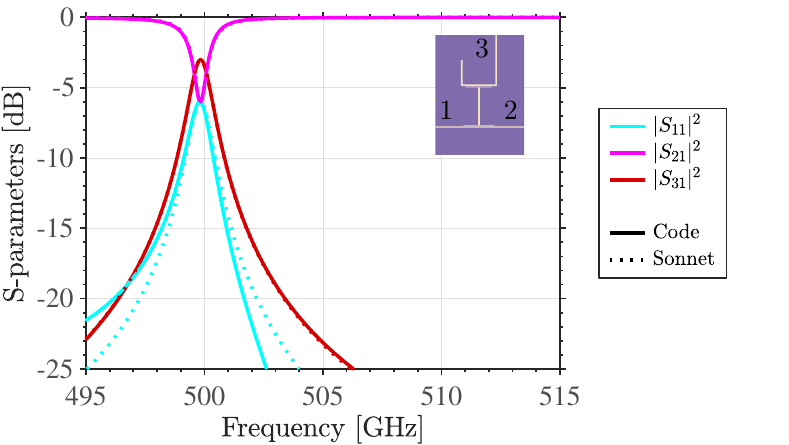}
			\caption{Comparison of the simulated S-parameters of 1 filter using Sonnet \cite{Sonnet} (resonator of \cref{fig:comparisonResonators}(c)) and the circuit model of \cref{fig:circuit_and_resonator_pictures}(a).}
			\label{fig:1filter_SonnetVScircuitModel}
		\end{figure}

\begin{figure}[t!]
			\centering
			\includegraphics{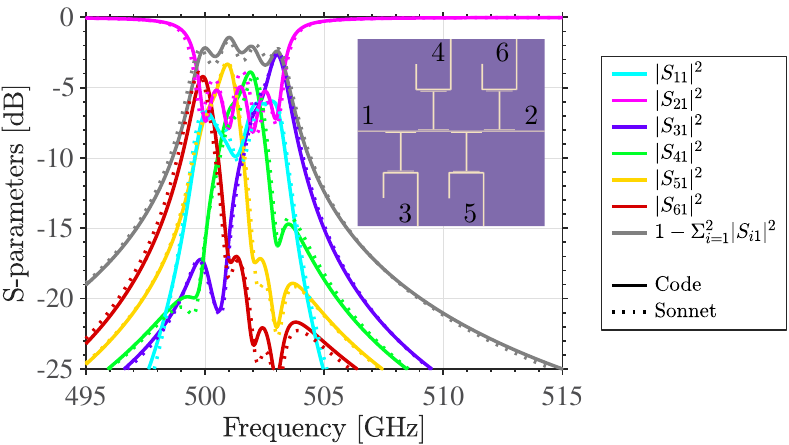}
			\caption{Comparison of the simulated S-parameters of 4 contiguous filters separated by $\lambda_{\mathrm{thru},i}/4$ using Sonnet \cite{Sonnet} and the circuit model of \cref{fig:circuit_and_resonator_pictures}(a).}
			\label{fig:4filters_SonnetVScircuitModel}
		\end{figure}

		The filter inside the dotted cyan box in \cref{fig:IshapedFilter} is simulated in Sonnet \cite{Sonnet}, and its performance is compared against the circuit model described in \cref{sec:circuitModel}. The comparison in \cref{fig:1filter_SonnetVScircuitModel} shows an excellent agreement between the performance of an I-shaped filter in isolation and the circuit model, with a slight deviation for the reflections off-resonance. We then investigated the arraying of 4 contiguous filters with an inter-filter separation $L_{\mathrm{thru},i}=\lambda_{\mathrm{thru},i}/4$. The agreement between the model and the full-wave simulations was found to be excellent in \cref{fig:4filters_SonnetVScircuitModel}; with a large difference in computation time between Sonnet \cite{Sonnet}, taking approximately 2 hours, and the model coded in Matlab \cite{Matlab}, lasting less than 0.5 seconds, using the same computer. This result is also relevant for claiming that our filters have a very low level of indirect cross-talk given that our circuit model can represent well their interactions with a transmission line model. With these results we were confident that we could use this code for predicting the behaviour of a large filter-bank with hundreds of channels (like depicted in \cref{fig:fullFilterBankResponse}), which would be otherwise a prohibitively heavy simulation.
		

	\section{Sparse Filter-Bank Chip}
	\label{sec:sparseFB}
		
		In this section we describe and characterize a spectrally sparse filter-bank consisting of 11 I-shaped filters in the band \SIrange{300}{400}{\giga\hertz} as schematically shown in \cref{fig:chip_cartoon}. Each THz filter pass-band is separated from its neighboring channels by 10 GHz in order to disentangle their response and get a full spectral characterization of their performance in semi-isolation. Each spectral channel (each filter-MKID pair) is physically separated from the next down the through-line by $L_{\mathrm{thru},i}=7\lambda_{\mathrm{thru},i}/4$. The targeted spectral resolution of the 11 filters was $Q_l=500$, which would require $Q_c=Q_{c1}=Q_{c2}\approx1179$ to maximize the coupling efficiency for $Q_i=3300$ following \cref{eq:optimumQc}. However, because previous experiments had consistently shown an increase of $Q_c$ with respect to the design value, for this chip we took the ansatz of designing for a lower coupling quality factor of $Q_c=500$. In order to decrease the $Q_c$ of the filters, while preserving their dimensions and the location of their THz resonances, the a\nobreakdash-Si dielectric thickness was increased from \SI{300}{\nano\meter} to \SI{370}{\nano\meter} and the sheet kinetic inductance $L_k$ of the microstrip top layer from \SI{1}{\pico\henry\per\sq} to \SI{1.17}{\pico\henry\per\sq}.
		
		\subsection{Fabrication}
		
\begin{figure}[t!]
			\centering
			\includegraphics[width=0.7\columnwidth]{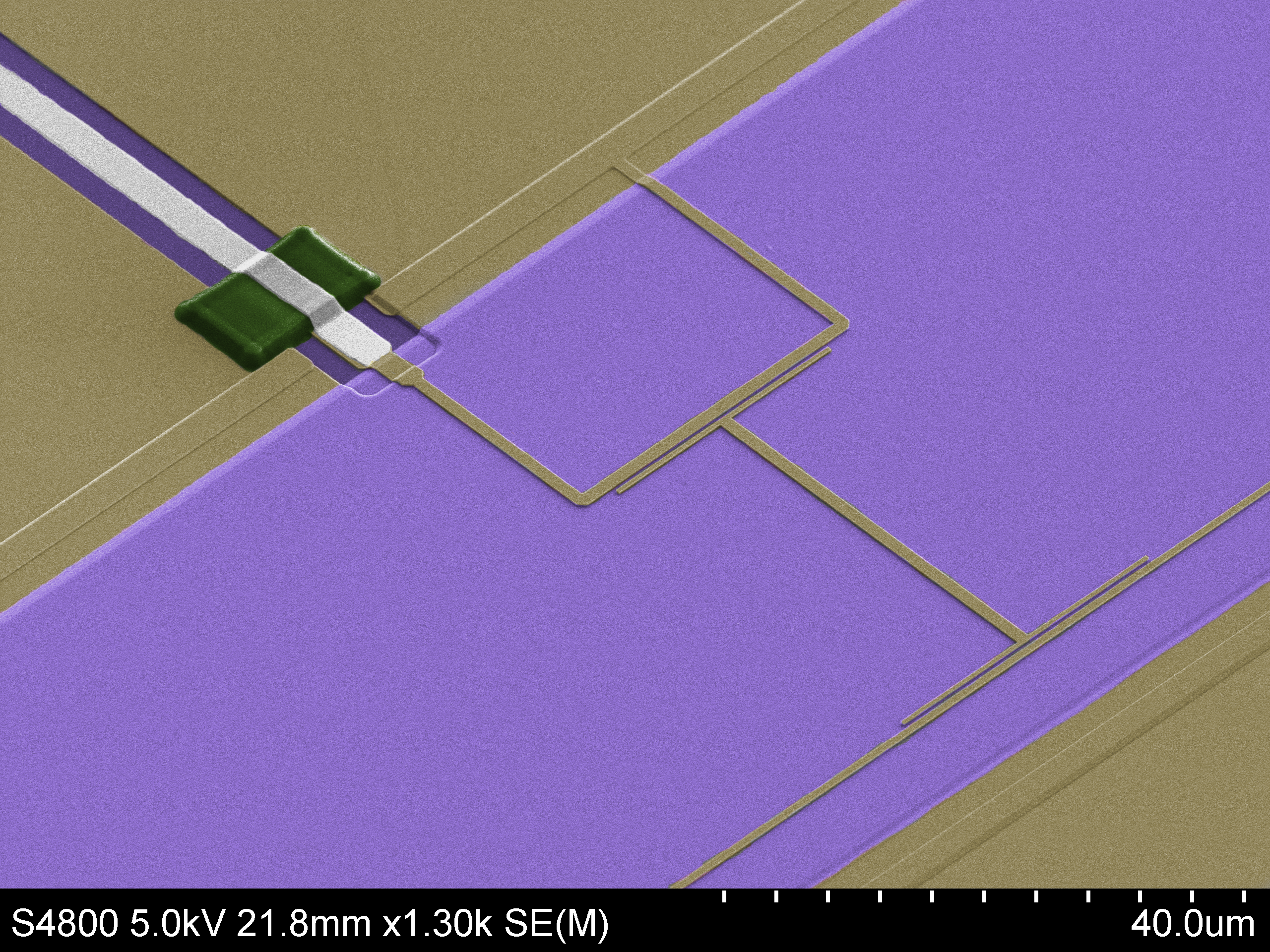}
			\caption{Colored SEM micrograph of a fabricated THz filter with an MKID on the top left of the image and the through-line in the bottom right.}
			\label{fig:SEM}
		\end{figure}

		The fabrication route of this chip is similar to \cite{microstripFP} and it is described in great detail in \cite{DESHIMA2_fab}. It starts with a \SI{260}{\nano\meter}-thick NbTiN layer ($T_c=\SI{15}{\kelvin}$, $\rho=\SI{90.0}{\micro\ohm\centi\meter}$) being deposited on top of a Si wafer using reactive sputtering of a NbTi target in a Nitrogen-Argon atmosphere \cite{uniformityNbTiN}. This layer is then patterned using photo-lithography with a positive resist and etched to define the microstrip filter-bank ground-plane and the central conductor of the CPW readout line. Afterwards, a \SI{370}{\nano\meter}-thick a-Si layer is deposited at \SI{250}{\celsius} using PECVD \cite{Bruno} and patterned to provide the dielectric support for the microstrips of the filter-bank. The microstrip lines are defined in a second NbTiN layer ($T_c=\SI{15}{\kelvin}$, $R_s=\SI{12.29}{\ohm\per\sq}$\footnote{Estimate using a four-probe resistance measurement at the edge of the wafer and the typical spatial variations of a film deposited with a small NbTi target in a Nordiko 2000 sputtering machine \cite{uniformityNbTiN}.} on a-Si) \SI{113}{\nano\meter} thick using a mix-and-match recipe employing a negative resist with electron-beam lithography to accurately pattern the smallest parts of the chip (\eg filter-bank) and photo-lithography to define the coarser parts (\eg MKIDs, readout lines, antenna, etc.) in a single lithographic step. After the double exposure, the resist is developed and the pattern is etched into the NbTiN. Next, a \SI{1}{\micro\meter}-thick layer of polyimide is spin-coated on the wafer, cured and patterned to provide the support for the bridges balancing the grounds of the CPW in the GHz readout and also to provide a good electrical contact between the Al and the NbTiN at the MKIDs. The \SI{40}{\nano\meter}-thick Al ($T_c=\SI{1.25}{\kelvin}$, $\rho=\SI{1.9}{\micro\ohm\centi\meter}$) layer is laid next, defining the central conductor of the narrow CPW section of the MKIDs. Lastly, a \SI{40}{\nano\meter}-thick layer of $\beta$-phase Ta ($T_c=\SI{0.95}{\kelvin}$, $\rho=\SI{239}{\micro\ohm\centi\meter}$) is deposited on the backside as an absorbing mesh for stray-light control \cite{SWcontrol}. The colored scanning electron microscope (SEM) micrograph in \cref{fig:SEM} shows a THz filter of the fabricated chip.
		
		\subsection{Measurements}
		To characterize the fabricated device, the chip is placed inside a commercial Bluefors dilution refrigerator, where it is cooled to \SI{120}{\milli\kelvin}. The chip is placed directly facing the cryostat window through an infrared filter stack with approximately $60\%$ transmission \cite{SebastianCryostat} in the band \SIrange{100}{700}{\giga\hertz} to limit the thermal loading. Moreover, we employed a wire-grid polarizer and a cryogenic quasi-optical band-pass filter stack (\SIrange{338}{373}{\giga\hertz} half-maximum pass-band) to minimize stray radiation on the chip. From the outside, a continuous-wave THz photo-mixing source (Toptica Terabeam 1550) illuminates the chip with a tunable single THz frequency. By sweeping the frequency of the source while recording the phase response of the tones probing the MKIDs, the frequency response is obtained with a spectral resolution of $\sim50$ MHz.
		
		In the forthcoming study we will experimentally evaluate all the relevant filter parameters, anticipating that the coupling strength of the wideband couplers $|S_{31}^\mathrm{wb}|^2$, which is needed to normalize the coupling efficiency of the filters ($|S_{i1}(f)|^2$), is not known with high accuracy due to fabrication tolerances and three-dimensional effects which are difficult to simulate. Therefore, we use the strategy of freeing the parameter $|S_{31}^\mathrm{wb}|^2$ and combining the peak and dip analyses to get the best consistent estimate in terms of coupling efficiency of the filters.
		
		In the first analysis we measure the transmission through the filter-bank, $|S_{21}(f)|^2$, which may be estimated from the ratio between the averaged phase response of the three wideband-coupled MKIDs after and before the filter-bank (respectively $\langle{}R_\mathrm{wb}^\mathrm{af}(f)\rangle{}$) and $\langle{}R_\mathrm{wb}^\mathrm{bf}(f)\rangle{}$) as
		\begin{equation}
			|S_{21}(f)|^2\approx\frac{\langle{}R_\mathrm{wb}^\mathrm{af}(f)\rangle{}}{\langle{}R_\mathrm{wb}^\mathrm{bf}(f)\rangle{}}.
			\label{eq:pseudoS21}
		\end{equation}
		
\begin{figure}[t!]
			\centering
			\includegraphics{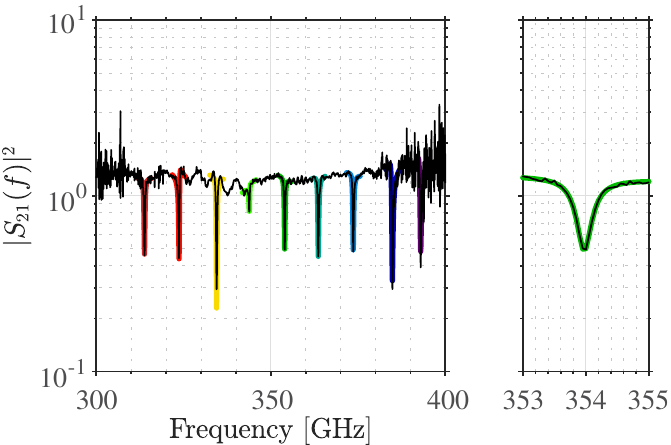}
			\caption{Pseudo transmission coefficient through the filter-bank estimated from \cref{eq:pseudoS21}, which is the averaged frequency response of the three wideband-coupled MKIDs trailing the filter-bank divided by that of the three leading it. The right-hand side panel emphasizes one of the skewed Lorentzian fits to the data with \cref{eq:skewedLorentzian}.}
			\label{fig:freqResponseMeasurements_dips}
		\end{figure}

		The dips emerging in \cref{fig:freqResponseMeasurements_dips} correspond to the filters extracting power from the through-line. Both the resonant frequency $f_0$ and the loaded quality factor $Q_l$ may be directly obtained by fitting the resonances of the pseudo transmission coefficient in \cref{fig:freqResponseMeasurements_dips} using a skewed Lorentzian function
		\begin{equation}
			L(f)=A_1\left(1+A_2(f-f_0)+\frac{A_3+A_4(f-f_0)}{1+4Q_l^2\frac{f-f_0}{f_0}}\right),
			\label{eq:skewedLorentzian}
		\end{equation}
		which follows from \cref{eq:S21_around_f0}. From the estimation of the resonant frequency $f_0$, the loaded quality factor $Q_l$ and the transmission dip depth $|S_{21}(f_0)|=\sqrt{L(f_0)/A_1}$, \cref{eq:S21_f0} allows the retrieval of $q_i$. Then, by means of \cref{eq:Ql_2port}, $Q_{c1}$ can be retrieved. An estimate of the coupling strength of the filters may be obtained at this point assuming $Q_c=Q_{c1}=Q_{c2}$ in \cref{eq:S31_f0,eq:qi}, which simplifies \cref{eq:S31_f0} to
		\begin{equation}
			|S_{31}(f_0)|^2=\frac{2}{\left(2+\frac{Q_c-q_i}{q_i}\right)^2}\approx|S_{i1}(f_i)|^2.
			\label{eq:pseudoS31_fromDip}
		\end{equation}

		Using the assumption $Q_{c1}=Q_{c2}$, $Q_i$ may be calculated from \cref{eq:qi}. This analysis on the dips of the $|S_{21}(f)|^2$ estimate of \cref{fig:freqResponseMeasurements_dips} results in the following average values: $\langle{}Q_l\rangle{}\approx960$, $\langle{}Q_c\rangle{}\approx2860$, $\langle{}|S_{i1}(f_i)|^2\rangle{}\approx-5.7$ dB and $\langle{}Q_i\rangle{}\approx2890$.
		
		In the next analysis we measure the coupling efficiency of each filter $|S_{i1}(f)|^2$, which may be estimated with
		\begin{equation}
			|S_{i1}(f)|^2\approx{}\frac{R_{\mathrm{filter},i}(f)}{\langle{}R_\mathrm{wb}^\mathrm{bf}(f)\rangle{}}|S_{31}^\mathrm{wb}|^2,
			\label{eq:pseudoS31}
		\end{equation}
		where $R_{\mathrm{filter},i}(f)$ is the phase response of the MKID attached to the $i^\mathrm{th}$ filter. The underlying assumptions for these estimates are that all the MKIDs have the same responsivity and that the filter-bank does not reflect a lot of power in-band. By fitting the peaks with the prototype function of \cref{eq:S31_around_f0}, the peak efficiency and the loaded quality factor can also be obtained. Subsequently using the knowledge of $Q_l$, $q_i$, $Q_{c1}$ from the dip analysis and the peak coupling efficiency of the filters $|S_{i1}(f_i)|^2$, the parameter $Q_{c2}$ may be obtained from \cref{eq:S31_f0} as
		\begin{equation}
			Q_{c2}=\frac{2Q_{c1}}{(1+Q_{c1}/Q_i)^2|S_{i1}(f_i)|^2}.
			\label{eq:Qc2}
		\end{equation}
		The three-port network internal quality factor $Q_i$ can be obtained henceforth by means of \cref{eq:qi}.
		
\begin{figure}[t!]
			\centering
			\includegraphics{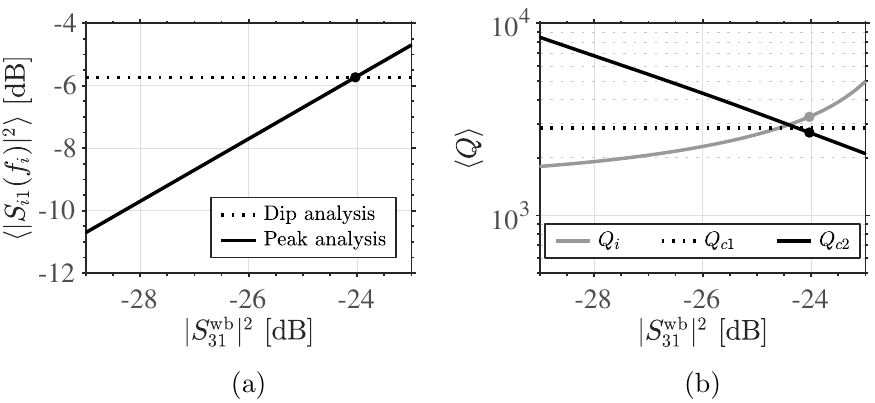}
			\caption{Sub-figure (a) is a consistency check between the peak and dip analyses for the average fitted coupling efficiency of the filters $\langle|S_{i1}(f_i)|^2\rangle$ using the strength of the wideband couplers $|S_{31}^\mathrm{wb}|^2$ as a free parameter. The value of $|S_{31}^\mathrm{wb}|^2\approx\SI{-24}{\decibel}$ (dot) makes the peak and dip analyses agree. Sub-figure (b) shows the quality factors $Q_i$, $Q_{c1}$ and $Q_{c2}$ as a function of the strength of the wideband coupler.}
			\label{fig.consistencyPeakvsDip}
		\end{figure}

		The problem with the peak analysis using \cref{eq:pseudoS31} is that, although we can directly measure the MKID reponses $R_{\mathrm{filter},i}(f)$ and $R_\mathrm{wb}^\mathrm{bf}(f)$, the wideband coupling strength $|S_{31}^\mathrm{wb}|^2$ cannot be directly obtained. Hence we perform a consistency check between the dip and peak analyses. To do so, we overlay in \cref{fig.consistencyPeakvsDip}(a) the mean value of the coupling efficiencies $\langle{}|S_{i1}(f_i)|^2\rangle{}$, obtained from the dip analysis with \cref{eq:pseudoS31_fromDip}, and the one obtained from the peak analysis by letting $|S_{31}^\mathrm{wb}|^2$ change in \cref{eq:pseudoS31}. In order to avoid data from low signal-to-noise ratio (SNR) measurements, the filters analyzed are well-within the 320--370 GHz quasi-optical pass-band, which is determined by the lens antenna in combination with the quasi-optical band-pass filter stack. It is apparent that the value that makes these experiments self-consistent is $|S_{31}^\mathrm{wb}|^2\approx-24$ dB, which is also in agreement with the design value simulated in Sonnet \cite{Sonnet} for this structure. Moreover, the dispersion between the average quality factors is reduced with this value as shown in \cref{fig.consistencyPeakvsDip}(b). With this congruent combination between the peak and dip analyses, based on the uncertainty of the strength of the wideband coupler, the filters can now be fully and unequivocally characterized.
		
\begin{figure}[t!]
			\centering
			\includegraphics{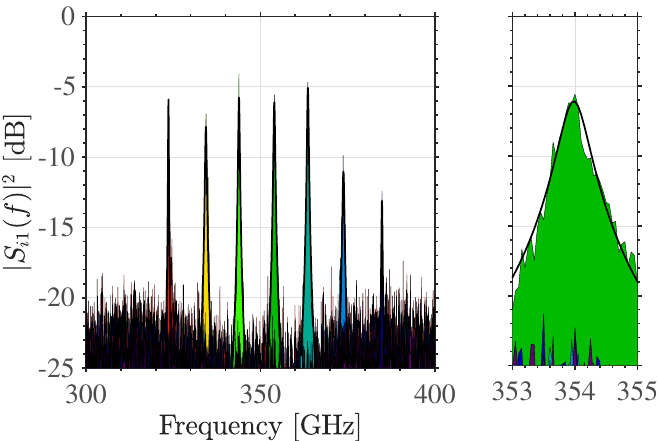}	
			\caption{Frequency response measurement of the pseudo coupling efficiency of the different filters estimated with \cref{eq:pseudoS31}. The right-hand side panel emphasizes one of the Lorentzian fits to the data. Only 7 out of 11 filters are visible above the noise floor due to the narrow quasi-optical pass-band determined by the antenna and the quasi-optical filter stack used.}
			\label{fig:freqResponseMeasurements_peaks}
		\end{figure}

\begin{figure}[t!]
			\centering
			\includegraphics{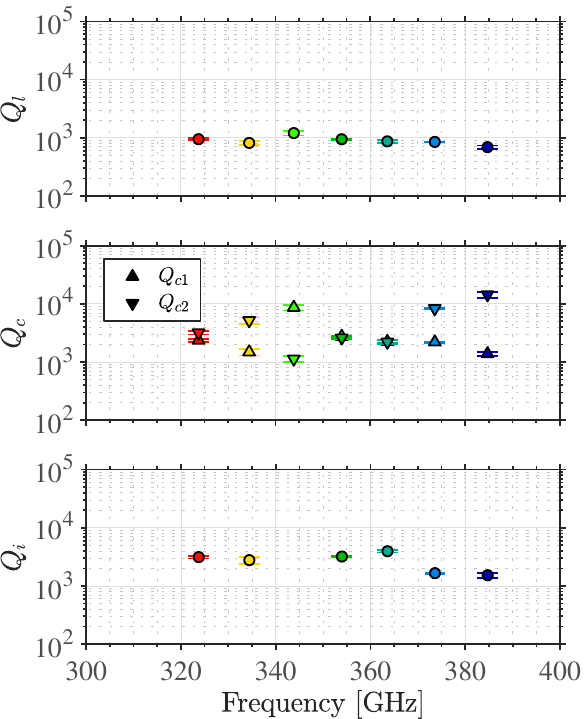}
			\caption{Measured quality factors of the filters using the peak and dip analyses.}
			\label{fig:QfactorsMeasurements}
		\end{figure}

		\Cref{fig:freqResponseMeasurements_peaks} depicts the coupling efficiency of the different band-pass filters as a function of frequency when estimated using \cref{eq:pseudoS31} with a coupling strength for the wideband-coupled MKID of $|S_{31}^\mathrm{wb}|^2\approx-24$ dB as inferred previously. By fitting the peaks with the prototype function of \cref{eq:S31_around_f0}, an average peak efficiency of $\langle|S_{i1}(f_i)|^2\rangle\approx-5.7$ dB can be obtained for those filters well-within the quasi-optical pass-band. \Cref{fig:QfactorsMeasurements} collects the different quality factors of these filters, whose average values are: $\langle{}Q_l\rangle{}\approx940$, which may be obtained from either the peak or the dip analyses; and $\langle{}Q_{c1}\rangle{}\approx2860$, $\langle{}Q_{c2}\rangle{}\approx2680$, and $\langle{}Q_i\rangle{}\approx3300$ which are acquired by combining the dip and the peak analyses.
		
		\subsection{Discussion}
		The yield of this chip has been 100\%, with 11 detectors observing 11 different THz pass-bands. The central frequencies of the filters have been observed to shift upwards by $\sim4\%$ from their designated values. This frequency upshift is likely due a the sheet kinetic inductance reduction in the top NbTiN layer caused by a lower resistivity than expected.
		
		The measured data for those filters well-within the quasi-optical pass-band show a modest peak coupling efficiency of $\langle|S_{i1}(f_i)|^2\rangle\approx-5.7$ dB and a loaded quality factor of $\langle{}Q_l\rangle{}\approx940$, which is higher than the targeted spectral resolution of 500. Both these results may be explained from the moderate internal quality factor $\langle{}Q_{i}\rangle{}\approx3300$ (which is slightly lower than the dielectric $Q_i$ reported in \cite{microstripFP}), and the high coupling quality factors $\langle{}Q_{c1}\rangle{}\approx2860$ and $\langle{}Q_{c2}\rangle{}\approx2680$ of the couplers surrounding the resonators.
		
\begin{figure}[t!]
			\centering
			\includegraphics{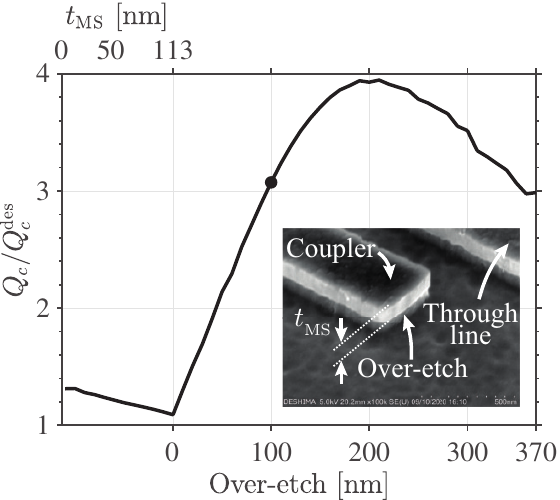}
			\caption{Simulated coupling quality factor variation as function of the microstrip metal thickness ($t_\mathrm{MS}$) and the a-Si over-etch using the fabricated dimensions of the filters, the measured electrical properties of the superconductors and the dielectric loss reported in \cite{microstripFP}. Instead $Q_c^\mathrm{des}$ is the design performance of the fully planar coupler using nominal dimensions and electrical properties on a lossless dielectric. The dot is the simulation with the closest approximation to the fabricated coupler shown in the micrograph of the inset.}
			\label{fig:overetch}
		\end{figure}

		In order to understand why $Q_c$ was too high with respect to the design value ($Q_c^\mathrm{des}=500$), we carefully inspected the filters using SEM. As can be seen in the inset of \cref{fig:overetch}, the a-Si was over-etched around \SI{100}{\nano\meter} after patterning the top NbTiN layer. The rest of the dimensions from the fabricated filters were altered by at most $5\%$. These fabrication tolerances have been analyzed in CST Microwave Studio \cite{CST} for a coupler in the through-line side of a filter nominally designed to operate at 350 GHz. $Q_c$ was found to strongly vary as a function of the dielectric over-etch as illustrated in \cref{fig:overetch}. With the \SI{100}{\nano\meter} over-etch measured we expect a threefold increase in $Q_c$ with respect to the design value, \ie $Q_c\approx3\cdot{}Q_c^\mathrm{des}=1500$. This qualitatively agrees with the observed increased quality factor, but is still quantitatively underestimating the measured $Q_c$.
		
		To partially compensate the impact of this fabrication tolerance, as well as for that of the thickness of the microstrips, we investigated the effect of re-depositing \SI{35}{\nano\meter} of a-Si on top of the filter-bank to reduce $Q_c$ by a factor 1.8 as simulated in CST Microwave Studio \cite{CST}. This addition has partially compensated the effect of these three-dimensional features, yielding $\langle{}Q_{c1}\rangle{}\approx1790$ and $\langle{}Q_{c2}\rangle{}\approx2010$, resulting in $\langle{}Q_l\rangle{}\approx640$ and a peak coupling strength $\langle{}|S_{i1}(f_i)|^2\rangle{}\approx-5.8$ dB. The loaded quality factor has been reduced as intended, however the coupling efficiency has not increased due to a lower $\langle{}Q_i\rangle{}\approx2140$, which is likely caused by higher dielectric losses in this \SI{35}{\nano\meter} a-Si layer (deposited at \SI{100}{\celsius}) \cite{Bruno}, which is also sensed by the microstrip fields.
		
		By properly tuning the coupling strength to ${Q_{c1}=Q_{c2}\approx1179}$, a loaded quality factor of ${Q_l\approx500}$ and a peak coupling efficiency of ${\langle|S_{i1}(f_i)|^2\rangle\approx\SI{-4.4}{\decibel}}$ should be within reach with the measured ${\langle{}Q_i\rangle{}\approx3300}$. Further improvement of the performance requires the development of low-loss and low-stress deposited dielectrics such as SiC. Moreover, to enhance the coupling efficiency beyond \SI{-3}{\decibel}, the spectrum could be over-sampled with more filters as described in \cite{SuperSpecCircuitModel}, however this would require an engineering effort to coherently or incoherently add the filtered signals at the detector level. Furthermore, sharper response tails could be achieved by adding extra filtering structures \cite{coupledFiltersTopHat,oversamplingFB,SuperSpecCircuitModel}, which can be beneficial to reduce cross-talk between neighboring channels.

	\section{Conclusions}
	\label{sec:conclusions}
		
		In this paper we propose a band-pass filter design suitable for mid-resolution broadband THz spectroscopy. The proposed filter geometry consists of an I-shaped microstrip half-wavelength resonator that couples energy with a spectral resolution $R=Q_l=f_0/\delta{}f$ from the through-line to the MKID sensing the spectral channel. From simulations, this resonator achieves on-resonance the optimal $-3$ dB coupling efficiency with a loaded quality factor of $Q_l\approx500$. Off-resonance, an extremely low disturbance is incurred by the filter to the through-line signal over a free spectral range of an octave, thus allowing for the arraying of many resonators in a large filter-bank configuration.
		
		A filter-bank chip sparsely sampling the \SIrange{300}{400}{\giga\hertz} spectrum with these I-shaped resonators has been fabricated in-house following a newly-developed micro-fabrication recipe. This fabrication route has shown a great level of reproducibility and accuracy, yielding a $100\%$ channel yield and achieving all the filter dimensions within $5\%$ error from their nominal values. The frequency response measurements of this chip showcases filters with a moderate spectral resolution of $\langle{}R\rangle{}=\langle{}Q_l\rangle{}\approx940$ with no spurious resonances in the band \SIrange{300}{400}{\giga\hertz}. The measured peak coupling efficiency averages $\langle|S_{i1}(f_i)|^2\rangle\approx-5.7$ dB $\approx27\%$, being currently limited by dielectric losses and amendable tolerances in the micro-fabrication such as the over-etch into the a-Si dielectric due to the patterning of the top NbTiN layer. This tolerance has largely increased the coupling quality factors $Q_{c1}$ and $Q_{c2}$ of the filters, which in turn has given rise to an exceedingly large average loaded quality factor at the cost of a moderate coupling efficiency. To amend the performance of the filters from this fabrication tolerance, a dielectric patch has been re-deposited on top of the filters to attempt the recovery of the intended coupling quality factor.

	\section*{Acknowledgment}
	The authors would like to thank Prof. Andrea Neto and Prof. Angelo Freni for the helpful discussions, and Sebastian H\"{a}hnle for easing the automation of the mask-making for our filter-bank chips.

	\ifCLASSOPTIONcaptionsoff
	\newpage
	\fi

	
	
	\bibliography{IEEEabrv,bibliography}

	\begin{thebibliography}{10}
		\providecommand{\url}[1]{#1}
		\csname url@samestyle\endcsname
		\providecommand{\newblock}{\relax}
		\providecommand{\bibinfo}[2]{#2}
		\providecommand{\BIBentrySTDinterwordspacing}{\spaceskip=0pt\relax}
		\providecommand{\BIBentryALTinterwordstretchfactor}{4}
		\providecommand{\BIBentryALTinterwordspacing}{\spaceskip=\fontdimen2\font plus
			\BIBentryALTinterwordstretchfactor\fontdimen3\font minus
			\fontdimen4\font\relax}
		\providecommand{\BIBforeignlanguage}[2]{{%
				\expandafter\ifx\csname l@#1\endcsname\relax
				\typeout{** WARNING: IEEEtran.bst: No hyphenation pattern has been}%
				\typeout{** loaded for the language `#1'. Using the pattern for}%
				\typeout{** the default language instead.}%
				\else
				\language=\csname l@#1\endcsname
				\fi
				#2}}
		\providecommand{\BIBdecl}{\relax}
		\BIBdecl
		
		\bibitem{highZgalaxies}
		C.~Carilli and F.~Walter, ``Cool gas in high-redshift galaxies,'' \emph{Annu.
			Rev. Astron. Astrophys.}, vol.~51, no.~1, pp. 105--161, Aug. 2013.
		
		\bibitem{submmGalaxies}
		A.~W. Blain, I.~Smail, R.~Ivison, J.-P. Kneib, and D.~T. Frayer,
		``Submillimeter galaxies,'' \emph{Phys. Rep.}, vol. 369, no.~2, pp. 111--176,
		Oct. 2002.
		
		\bibitem{THz_astronomy}
		C.~{Kulesa}, ``Terahertz spectroscopy for astronomy: From comets to
		cosmology,'' \emph{{IEEE} Trans. {THz} Sci. Technol.}, vol.~1, no.~1, pp.
		232--240, Sep. 2011.
		
		\bibitem{CIB}
		H.~Dole \emph{et~al.}, ``The cosmic infrared background resolved by {S}pitzer -
		{C}ontributions of mid-infrared galaxies to the far-infrared background,''
		\emph{Astron. Astrophys.}, vol. 451, no.~2, pp. 417--429, May 2006.
		
		\bibitem{DESHIMA1}
		A.~Endo \emph{et~al.}, ``Wideband on-chip terahertz spectrometer based on a
		superconducting filterbank,'' \emph{J. Astron. Telesc. Instrum. Syst.},
		vol.~5, no.~3, pp. 1--12, Jun. 2019.
		
		\bibitem{SuperSpec}
		J.~Redford \emph{et~al.}, ``{The design and characterization of a 300 channel,
			optimized full-band millimeter filterbank for science with SuperSpec},'' in
		\emph{Proc. {SPIE}, Millimeter, Submillimeter, and Far-Infrared Detectors and
			Inst. for Astron. IX}, J.~Zmuidzinas and J.-R. Gao, Eds., vol. 10708, Int.
		Soc. Opt. Photon.\hskip 1em plus 0.5em minus 0.4em\relax SPIE, Jul. 2018, pp.
		187--194.
		
		\bibitem{CAMELS}
		C.~N. Thomas \emph{et~al.}, ``Progress on the cambridge emission line surveyor
		({CAMELS}),'' in \emph{Proc. 26th Int. Symp. Space THz Technol.}, Mar. 2015.
		
		\bibitem{IFU}
		\BIBentryALTinterwordspacing
		{European Southern Observatory (ESO)}. Integral field units. Accessed:
		2021-02-08. [Online]. Available:
		\url{https://www.eso.org/public/teles-instr/technology/ifu/}
		\BIBentrySTDinterwordspacing
		
		\bibitem{3dsky_1}
		J.~Geach \emph{et~al.}, ``The case for a {'sub-millimeter SDSS'}: a {3D} map of
		galaxy evolution to z$\sim$10,'' \emph{Bull. Am. Astron. Soc.}, vol.~51,
		no.~3, May 2019.
		
		\bibitem{3dsky_2}
		D.~Obreschkow, H.-R. Klöckner, I.~Heywood, F.~Levrier, and S.~Rawlings, ``{A
			Virtual sky with extragalactic HI and CO lines for the Square Kilometre Array
			and the Atacama Large Millimeter/Submillimeter Array},'' \emph{Astrophys.
			J.}, vol. 703, no.~2, pp. 1890--1903, Sep. 2009.
		
		\bibitem{SZ_astrophysics}
		T.~Mroczkowski \emph{et~al.}, ``Astrophysics with the spatially and spectrally
		resolved {Sunyaev-Zeldovich} effects,'' \emph{Space Sci. Rev.}, vol. 215,
		no.~1, p.~17, Feb. 2019.
		
		\bibitem{DESHIMA1_nature}
		A.~Endo \emph{et~al.}, ``First light demonstration of the integrated
		superconducting spectrometer,'' \emph{Nat. Astron.}, vol.~3, no.~11, pp.
		989--996, Nov. 2019.
		
		\bibitem{SuperSpecCircuitModel}
		A.~Kov{\'a}cs \emph{et~al.}, ``Superspec: design concept and circuit
		simulations,'' in \emph{Proc. {SPIE}, Millimeter, Submillimeter, and
			Far-Infrared Detectors and Instrumentation for Astronomy VI}, J.~Z. Wayne
		S.~Holland, Ed., vol. 8452, Int. Soc. Opt. Photon.\hskip 1em plus 0.5em minus
		0.4em\relax SPIE, Jul. 2012, pp. 748--757.
		
		\bibitem{GeorgeChePhD}
		G.~{Che}, ``Advancements in kinetic inductance detector, spectrometer, and
		amplifier technologies for millimeter-wave astronomy,'' Ph.D. dissertation,
		Arizona State University, 2018.
		
		\bibitem{radiationLossCPW}
		S.~H{\"a}hnle \emph{et~al.}, ``Suppression of radiation loss in high kinetic
		inductance superconducting co-planar waveguides,'' \emph{Appl. Phys. Lett.},
		vol. 116, no.~18, p. 182601, May 2020.
		
		\bibitem{MKID}
		P.~K. Day, H.~G. LeDuc, B.~A. Mazin, A.~Vayonakis, and J.~Zmuidzinas, ``A
		broadband superconducting detector suitable for use in large arrays,''
		\emph{Nature}, vol. 425, no. 6960, pp. 817--821, Oct. 2003.
		
		\bibitem{hybridMKID}
		R.~M.~J. Janssen \emph{et~al.}, ``High optical efficiency and photon noise
		limited sensitivity of microwave kinetic inductance detectors using phase
		readout,'' \emph{Appl. Phys. Lett.}, vol. 103, no.~20, p. 203503, Nov. 2013.
		
		\bibitem{KIDreadout_Baselmans}
		J.~{van Rantwijk}, M.~{Grim}, D.~{van Loon}, S.~{Yates}, A.~{Baryshev}, and
		J.~{Baselmans}, ``Multiplexed readout for 1000-pixel arrays of microwave
		kinetic inductance detectors,'' \emph{{IEEE} Trans. Microw. Theory Tech.},
		vol.~64, no.~6, pp. 1876--1883, Jun. 2016.
		
		\bibitem{KIDreadout_Mazin}
		S.~McHugh \emph{et~al.}, ``A readout for large arrays of microwave kinetic
		inductance detectors,'' \emph{Rev. Sci. Instrum.}, vol.~83, no.~4, p. 044702,
		Apr. 2012.
		
		\bibitem{coupledFiltersTopHat}
		A.~Endo \emph{et~al.}, ``Design of an integrated filterbank for {DESHIMA}:
		On-chip submillimeter imaging spectrograph based on superconducting
		resonators,'' \emph{J. Low Temp. Phys.}, vol. 167, no.~3, pp. 341--346, Jan.
		2012.
		
		\bibitem{oversamplingFB}
		E.~Shirokoff \emph{et~al.}, ``Mkid development for {SuperSpec}: an on-chip,
		mm-wave, filter-bank spectrometer,'' in \emph{Proc. {SPIE}, Millimeter,
			Submillimeter, and Far-Infrared Detectors and Inst. for Astron. VI}, J.~Z.
		Wayne S.~Holland, Ed., vol. 8452, Int. Soc. Opt. Photon.\hskip 1em plus 0.5em
		minus 0.4em\relax SPIE, Jul. 2012, pp. 209--219.
		
		\bibitem{cochlea}
		C.~J. {Galbraith}, R.~D. {White}, L.~{Cheng}, K.~{Grosh}, and G.~M. {Rebeiz},
		``Cochlea-based {RF} channelizing filters,'' \emph{{IEEE} Trans. Circuits
			Syst. I}, vol.~55, no.~4, pp. 969--979, May 2008.
		
		\bibitem{WSPEC}
		S.~Bryan \emph{et~al.}, ``{WSPEC:} a waveguide filter-bank focal plane array
		spectrometer for millimeter wave astronomy and cosmology,'' \emph{J. Low
			Temp. Phys.}, vol. 184, no.~1, pp. 114--122, Jul. 2016.
		
		\bibitem{Matthei_decouplingResonators}
		G.~Matthaei, L.~Young, and E.~Jones, \emph{Microwave Filters,
			Impedance-matching Networks, and Coupling Structures}, ser. Artech House
		Microwave Library.\hskip 1em plus 0.5em minus 0.4em\relax Norwood, MA: Artech
		House Books, 1980, ch.~16, p. 973.
		
		\bibitem{BenMazinPhD}
		B.~A. Mazin, ``Microwave kinetic inductance detectors,'' Ph.D. dissertation,
		California Institute of Technology, 2005.
		
		\bibitem{Akira_APL}
		A.~Endo \emph{et~al.}, ``On-chip filter bank spectroscopy at 600--700 {GHz}
		using {NbTiN} superconducting resonators,'' \emph{Appl. Phys. Lett.}, vol.
		103, no.~3, p. 032601, Jul. 2013.
		
		\bibitem{RamiBarendsPhD}
		R.~Barends, ``Photon-detecting superconducting resonators,'' Ph.D.
		dissertation, Delft University of Technology, 2009.
		
		\bibitem{Frickey}
		D.~A. {Frickey}, ``{Conversions between S, Z, Y, H, ABCD, and T parameters
			which are valid for complex source and load impedances},'' \emph{{IEEE}
			Trans. Microw. Theory Tech.}, vol.~42, no.~2, pp. 205--211, Feb. 1994.
		
		\bibitem{MBtheory}
		D.~C. Mattis and J.~Bardeen, ``Theory of the anomalous skin effect in normal
		and superconducting metals,'' \emph{Phys. Rev.}, vol. 111, pp. 412--417, Jul.
		1958.
		
		\bibitem{Bruno}
		B.~T. Buijtendorp \emph{et~al.}, ``Characterization of low-loss hydrogenated
		amorphous silicon ﬁlms for superconducting resonators,'' in \emph{Proc.
			{SPIE}, Millimeter, Submillimeter, and Far-Infrared Detectors and Inst. for
			Astron. X}, J.~Zmuidzinas and J.-R. Gao, Eds., vol. 11453, Int. Soc. Opt.
		Photon.\hskip 1em plus 0.5em minus 0.4em\relax SPIE, Dec. 2020, pp. 459--472.
		
		\bibitem{slotTwoMedia}
		A.~Neto and S.~Maci, ``{Green's function for an infinite slot printed between
			two homogeneous dielectrics---Part I: Magnetic currents},'' \emph{IEEE Trans.
			Antennas Propag.}, vol.~51, no.~7, pp. 1572--1581, Jul. 2003.
		
		\bibitem{TLTool}
		S.~van Berkel, A.~Garufo, N.~{Llombart Juan}, and A.~Neto,
		``\BIBforeignlanguage{English}{A quasi-analytical tool for the
			characterization of transmission lines at high frequencies [{EM} programmer's
			notebook]},'' \emph{\BIBforeignlanguage{English}{{IEEE} Antennas Propag.
				Mag.}}, vol.~58, no.~3, pp. 82--90, Jun. 2016.
		
		\bibitem{CPWmodes}
		M.~{Spirito}, G.~{Gentile}, and A.~{Akhnoukh}, ``Multimode analysis of
		transmission lines and substrates for (sub)mm-wave calibration,'' in
		\emph{82nd ARFTG Microw. Meas. Conf.}, Nov. 2013, pp. 1--6.
		
		\bibitem{CST}
		\BIBentryALTinterwordspacing
		{CST Microwave Studio (2018)}. {Dassault Systèmes Simulia Corp.} [Online].
		Available:
		\url{https://www.3ds.com/products-services/simulia/products/cst-studio-suite/}
		\BIBentrySTDinterwordspacing
		
		\bibitem{elbowCoupler}
		J.~Gao, J.~Zmuidzinas, B.~Mazin, P.~Day, and H.~Leduc, ``Experimental study of
		the kinetic inductance fraction of superconducting coplanar waveguide,''
		\emph{Nucl. Instrum. Methods Phys. Res. A}, vol. 559, no.~2, pp. 585--587,
		Apr. 2006, proc. 11th Int. Workshop on Low Temp. Detectors.
		
		\bibitem{microwaveLosses}
		A.~D. O’Connell \emph{et~al.}, ``Microwave dielectric loss at single photon
		energies and millikelvin temperatures,'' \emph{App. Phys. Lett.}, vol.~92,
		no.~11, p. 112903, Mar. 2008.
		
		\bibitem{Superspec_Qi}
		S.~Hailey-Dunsheath \emph{et~al.}, ``Optical measurements of {SuperSpec}: A
		millimeter-wave on-chip spectrometer,'' \emph{J. Low Temp. Phys.}, vol. 176,
		no.~5, pp. 841--847, Sep. 2014.
		
		\bibitem{microstripFP}
		S.~H{\"a}hnle \emph{et~al.}, ``Superconducting microstrip losses at microwave
		and sub-mm wavelengths,'' accepted for publication in \emph{Phys. Rev. Appl.}
		
		\bibitem{Sonnet}
		\BIBentryALTinterwordspacing
		{Sonnet em (17.56)}. {Sonnet Software, Inc.} [Online]. Available:
		\url{https://www.sonnetsoftware.com/}
		\BIBentrySTDinterwordspacing
		
		\bibitem{spuriousResRing}
		L.-H. Hsieh and K.~Chang, ``Simple analysis of the frequency modes for
		microstrip ring resonators of any general shape and correction of an error in
		the literature,'' \emph{Microw. Opt. Technol. Lett.}, vol.~38, no.~3, pp.
		209--213, Jun. 2003.
		
		\bibitem{TLS}
		J.~Gao \emph{et~al.}, ``A semiempirical model for two-level system noise in
		superconducting microresonators,'' \emph{Appl. Phys. Lett.}, vol.~92, no.~21,
		p. 212504, May 2008.
		
		\bibitem{Matlab}
		\BIBentryALTinterwordspacing
		{Matlab (R2020b)}. {The Mathworks, Inc.} [Online]. Available:
		\url{https://www.mathworks.com/products/matlab.html}
		\BIBentrySTDinterwordspacing
		
		\bibitem{DESHIMA2_fab}
		D.~Thoen, V.~Murugesan, K.~Karatsu, A.~{Pascual Laguna}, A.~Endo, and J.~J.~A.
		Baselmans, ``Combining {UV}-and electron-beam lithography for superconducting
		bandpass filters in mm/sub-mm astronomy,'' in \emph{Proc. {SPIE}, Millimeter,
			Submillimeter, and Far-Infrared Detectors and Instrumentation for Astronomy
			X}, J.~Zmuidzinas and J.-R. Gao, Eds., vol. 11453, Int. Soc. Opt.
		Photon.\hskip 1em plus 0.5em minus 0.4em\relax SPIE, Dec. 2020.
		
		\bibitem{uniformityNbTiN}
		D.~J. {Thoen}, B.~G.~C. {Bos}, E.~A.~F. {Haalebos}, T.~M. {Klapwijk}, J.~J.~A.
		{Baselmans}, and A.~{Endo}, ``Superconducting {NbTiN} thin films with highly
		uniform properties over a ${\varnothing}$ 100 mm wafer,'' \emph{{IEEE} Trans.
			Appl. Supercond.}, vol.~27, no.~4, pp. 1--5, Jun. 2017.
		
		\bibitem{SWcontrol}
		S.~J.~C. {Yates} \emph{et~al.}, ``Surface wave control for large arrays of
		microwave kinetic inductance detectors,'' \emph{{IEEE} Trans. {THz} Sci.
			Technol.}, vol.~7, no.~6, pp. 789--799, Nov. 2017.
		
		\bibitem{SebastianCryostat}
		S.~H{\"a}hnle, J.~Bueno, R.~Huiting, S.~J.~C. Yates, and J.~J.~A. Baselmans,
		``Large angle optical access in a sub-kelvin cryostat,'' \emph{J. Low Temp.
			Phys.}, vol. 193, no.~5, pp. 833--840, Dec. 2018.
		
	\end{thebibliography}
	%
	
\end{document}